\documentclass[runningheads]{llncs}
\usepackage{amsmath, ulem, graphicx, marvosym, fancyhdr, amscd, amssymb, mathrsfs,color,subcaption,nicefrac,epstopdf}
\usepackage{framed}
\usepackage{hyperref}

\begin{document} 

\title{High Definition image classification in Geoscience using Machine Learning}

%
%
\author{Yajun An\inst{1}\orcidID{0000-0002-9898-4882} \and
Zachary Golden\inst{2} \and
Tarka Wilcox\inst{3} \and 
Renzhi Cao\inst{2}\orcidID{0000-0002-8345-343X}}
\authorrunning{Y. An et al.}
%
\institute{University of Washington-Tacoma, Tacoma, WA, 98402  \and
Pacific Lutheran University, Tacoma, WA, 98447
\and
Wiss, Janney, Elstner Associates, Inc. Denver, CO, 80235
\email{\\yajuna@uw.edu; TWilcox@wje.com; caora@plu.edu}}
\maketitle              
\begin{abstract}
High Definition (HD) digital photos taken with drones are widely used in the study of Geoscience. However, blurry images are often taken in collected data, and it takes a lot of time and effort to distinguish clear images from blurry ones. In this work, we apply Machine learning techniques, such as Support Vector Machine (SVM) and Neural Network (NN) to classify HD images in Geoscience as clear and blurry, and therefore automate data cleaning in Geoscience. We compare the results of classification based on features abstracted from several mathematical models. Some of the implementation of our machine learning tool is freely available at: \url{ https://github.com/zachgolden/geoai}.

\end{abstract}

\section*{Introduction} Blur in Geoscience photos is a very common phenomenon, especially when taken by drones (Blur could come from defocus, camera shake, motion, etc.). Many photos are not useful due to blur even for HD photo. It is particularly challenging for data cleaning in Geoscience, even for a clear image. Other than conventional blur detection, we also need to be careful with false positive for blurs. Currently, it takes days at a time to clean data by picking out blurry images manually. In this work, we develop tools using machine learning techniques for data cleaning in Geoscience.


Methods have been proposed for blur detection based on the three strategies: Full-reference image quality assessment (FR-IQA) techniques that compare a reference and a test image and predict the perceptual quality of the test image in terms of a scalar value representing an objective score; No-reference image quality assessment (NR-IQA) techniques that measure the perceptual quality of an image without access to the reference image; and Reduced-reference image quality assessment (RR-IQA) techniques that provide a solution that lies between FR and NR models \cite{de_masilamani}. 

Mathematical models have been used for those blur detection methods. Hsu and Chen proposed a blur detector using support vector machine with image gradient model \cite{hsu_chen}; De and Masilamani proposed a image sharpness measure in the frequency domain using the Fast Fourier Transform \cite{de_masilamani}; Tong et al proposed wavelet transforms in blur detection \cite{tong_etal}. In this work, we present a survey of several mathematical models for feature analysis and construct several metrics for measuring blurriness of a given clear image to itself with added noise. We also explore machine learning technique for this problem since it has been successfully and widely applied to solve problems in different fields, including image recognition, bioinformatics, voice recognition, drug discovery, etc. \cite{lecun2015deep,lee2017deep,hou2019protein,staples2019artificial,stephenson2019survey}

\section{Mathematical models}
 The input data in this work is a {\it grayscale} image of size $M\times N$, which is represented as $M\times N$ pixels while each pixel contains an integer between 0 to 255. 

\subsection{Feature analysis: image gradient model}

We follow Hsu and Chen's work \cite{hsu_chen} and compute the gradient of an image at each pixel: let $F(i,j)$ denote the pixel of image $F$ at pixel $(i,j)$. Then the gradient at $(i,j)$ is the vector $\left<\frac{\partial F}{\partial x}, \frac{\partial F}{\partial y}\right>\bigg\rvert_{(i,j)}$. We will use the {\it gradient magnitude} and the {\it gradient direction} as features of image $F$:
\begin{equation}G=\sqrt{\left(\frac{\partial F}{\partial x}\right)^2+ \left(\frac{\partial F}{\partial y}\right)^2},\qquad \theta=\arctan\left(\frac{\nicefrac{\partial F}{\partial x}}{\nicefrac{\partial F}{\partial y}}\right).\end{equation}

In a clear image, shapes should have sharp edges and angles, therefore the difference between pixels would be ``big", producing a large gradient magnitude and small angle. Whereas blurry images will give blurry edges and rounded angles. We base our first feature on pixel count and feed the histogram of gradient information to the machine:

\begin{framed}
\noindent {\bf Input data}: Image $F$ of size $M\times N$

\noindent {\bf Output feature}: clear coefficient $\alpha$. 

\begin{enumerate}
\item Compute the gradient of image and store as a $M\times N$ array.
\item Compute the gradient magnitude and gradient direction, then convert the $M\times N$ arrays to two vector of length $MN$.
\item Compute the percentage of pixels with high gradient magnitude (with preset threshold) over all pixels, we call this the clear coefficient $\alpha$.
\end{enumerate}
\end{framed}

\begin{figure}
\centering
  \begin{tabular}{c @{\qquad} c }
    \includegraphics[width=.4\linewidth]{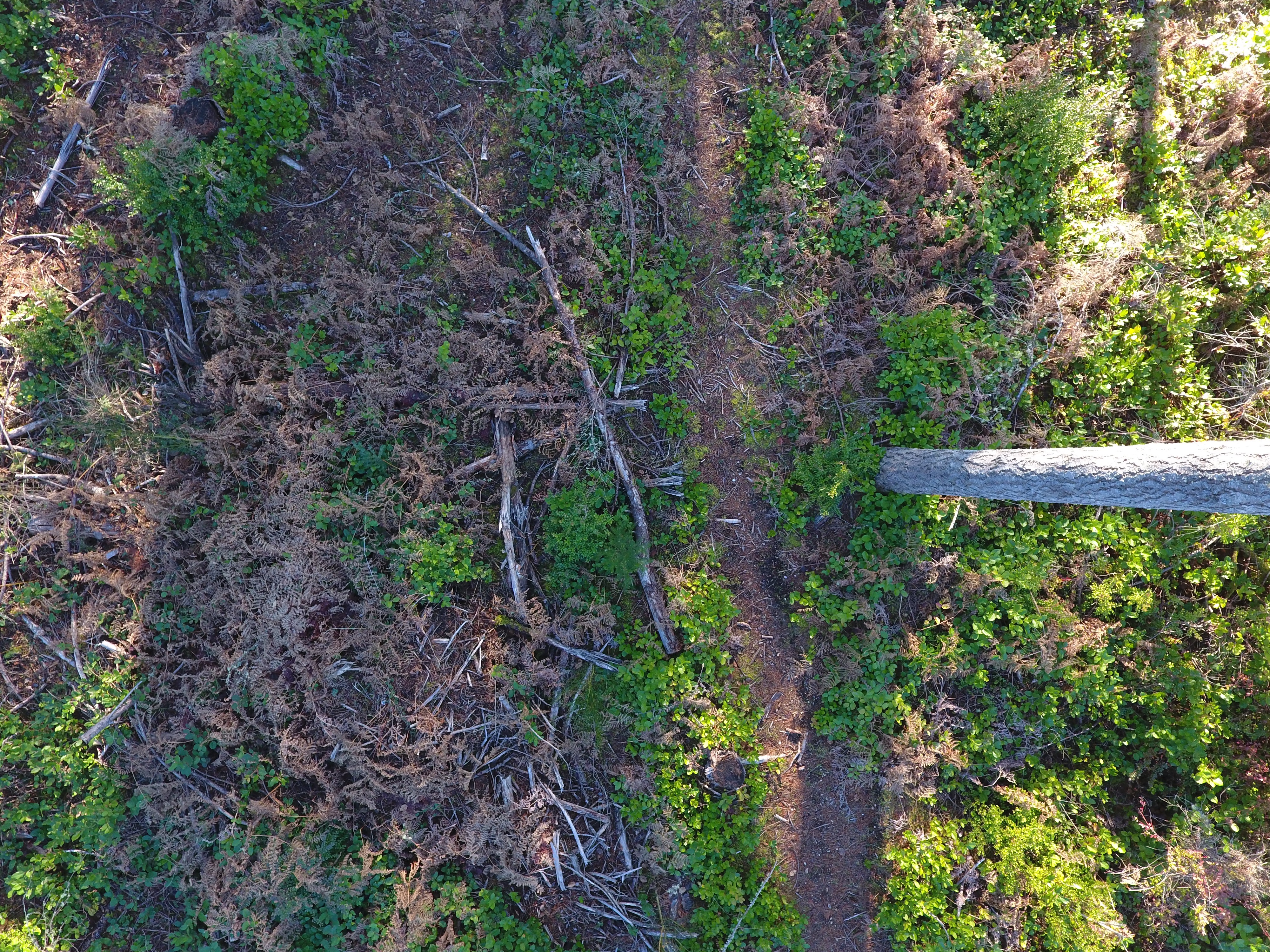} &
    \includegraphics[width=.4\linewidth]{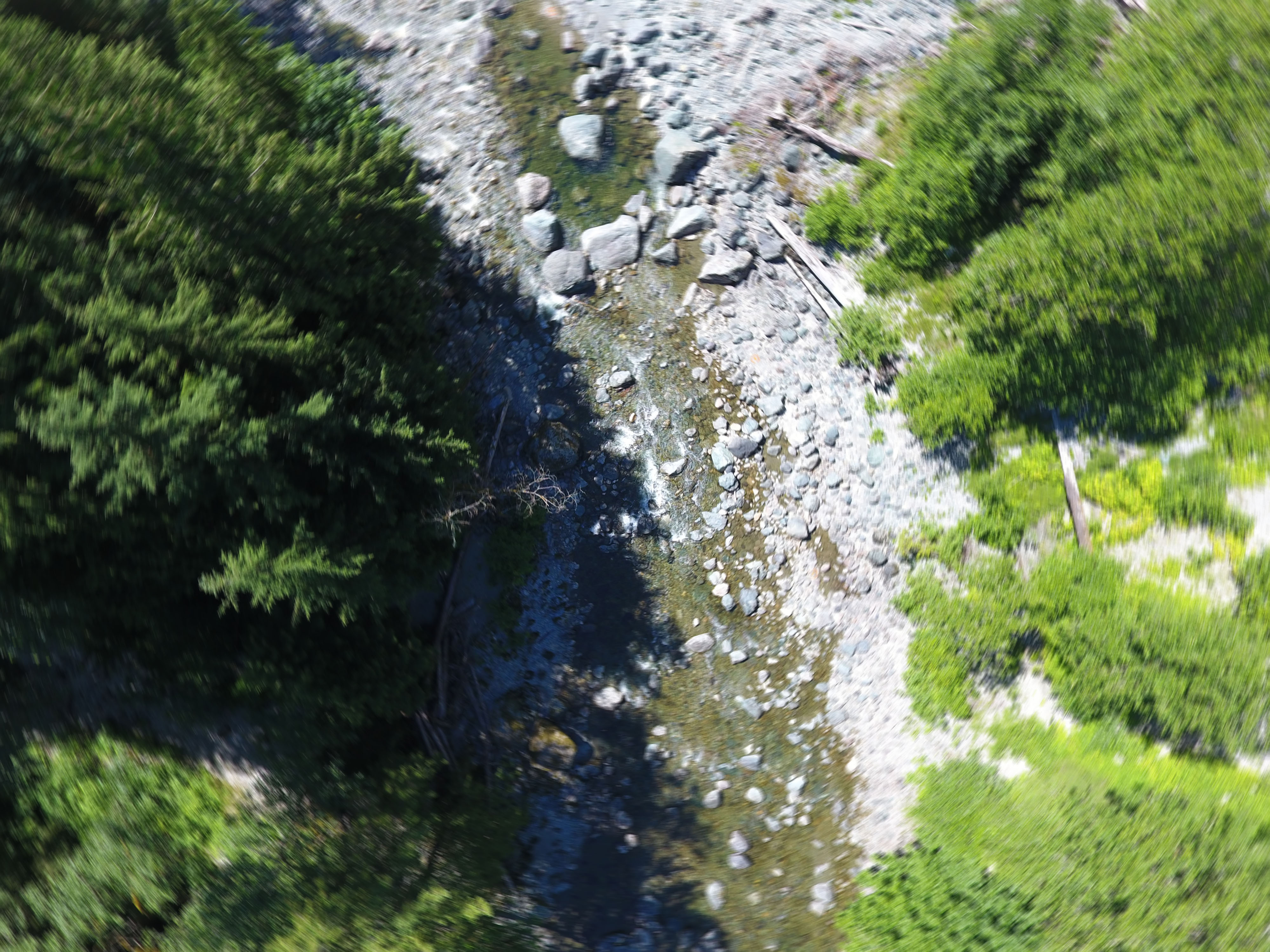} \\
    \small (a) Clear & \small (b) Spin 1
  \end{tabular}
\caption{Clear image, and generated blurry image with radial spin blur extent 1. Radial spin blur comes from the instability of drone imaging. }
\label{fig1}
\end{figure}
\begin{figure}
\centering
  \includegraphics[width=0.85\linewidth]{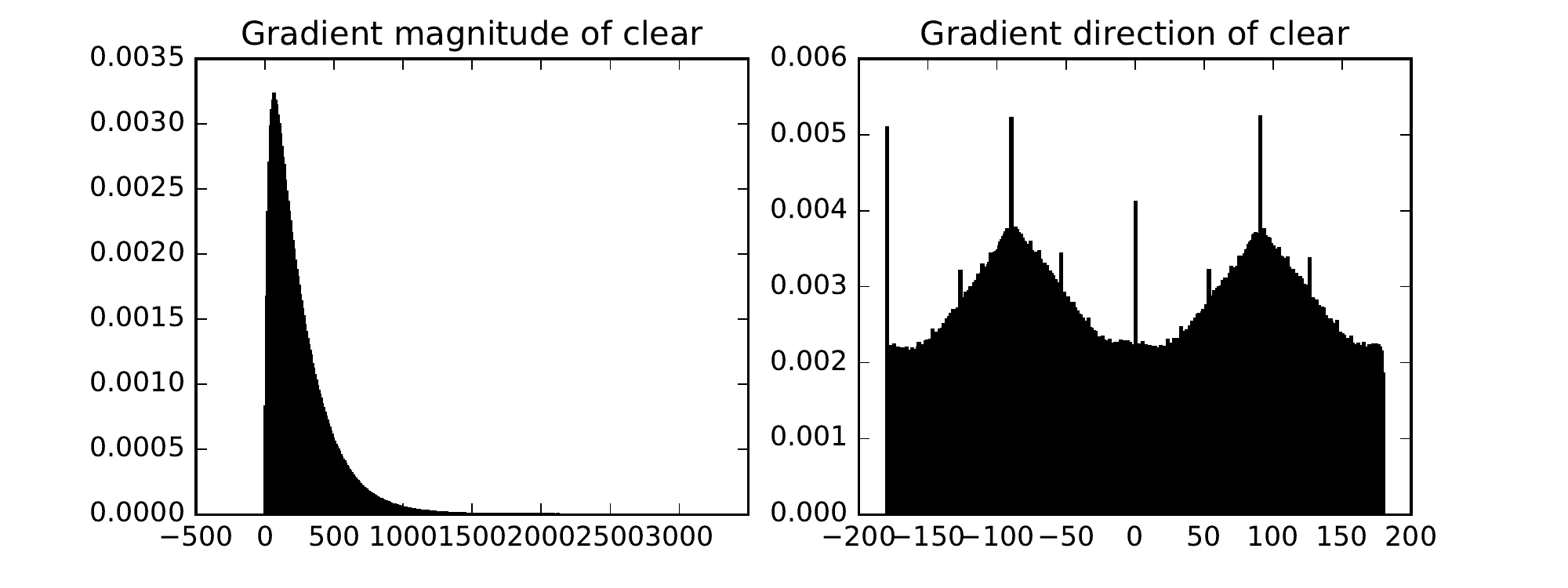}
  \caption{Histogram of gradient magnitude and gradient direction for clear image. Clear coefficient $\alpha$ is 0.0111}
  \label{fig:cleargradient}
\end{figure}
\begin{figure}
  \centering
  \includegraphics[width=0.85\linewidth]{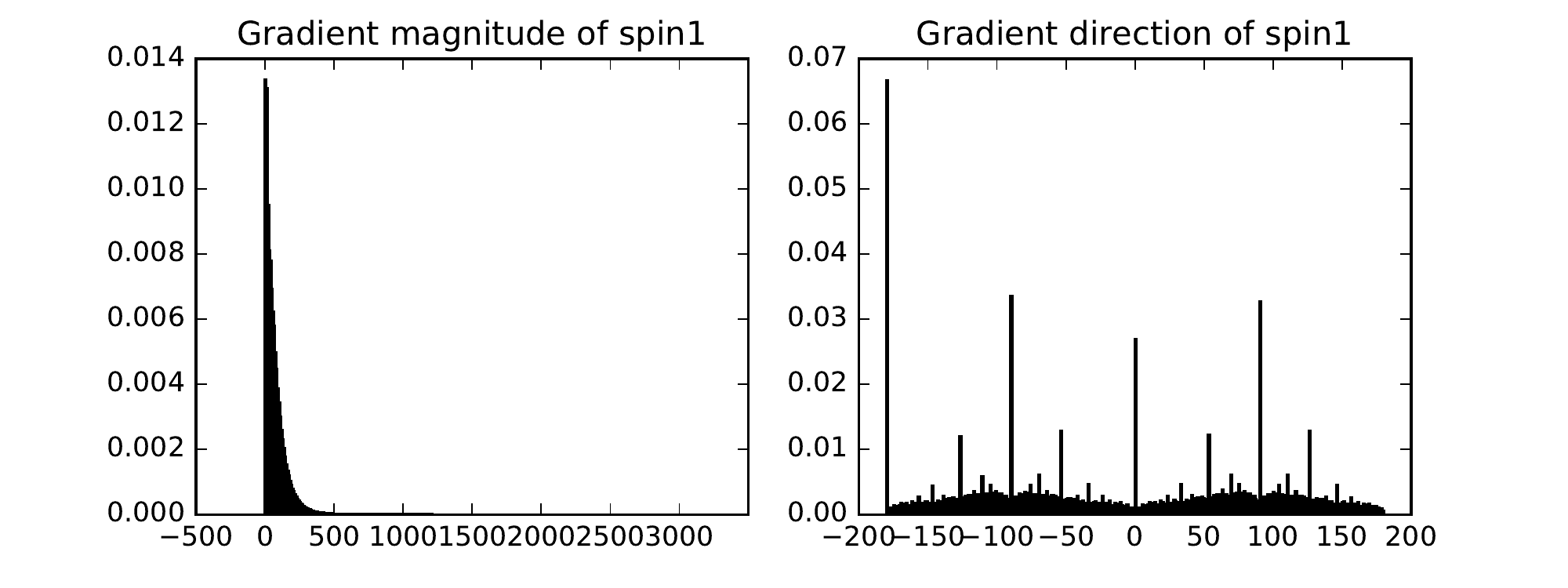}
  \caption{Histogram of gradient magnitude and gradient direction for spin 1 image. Clear coefficient $\alpha$ is 0.0001.}
  \label{fig:spingradient}
  
\end{figure}

\ref{fig1}(a) is classified manually as clear, and \ref{fig1}(b) is artificially blurred by spinning. Notice the two images have distinct characteristics, but our feature analysis in Figures \ref{fig:cleargradient} and \ref{fig:spingradient} shows a large number of pixels with gradient larger than 1000 in the clear image, and none in the blurry one. Hence our method is image independent. In the histogram of gradient direction, blurry image angle is a lot fuzzier with large spikes and general lower counts of pixels. These histograms can be used as features to distinguish clear and blurry images: we expect more blur to result in fewer counts of large magnitudes and larger values and variations of gradient direction.

\begin{figure}
\begin{subfigure}{0.5\textwidth}
  \centering
  \includegraphics[width=.8\linewidth]{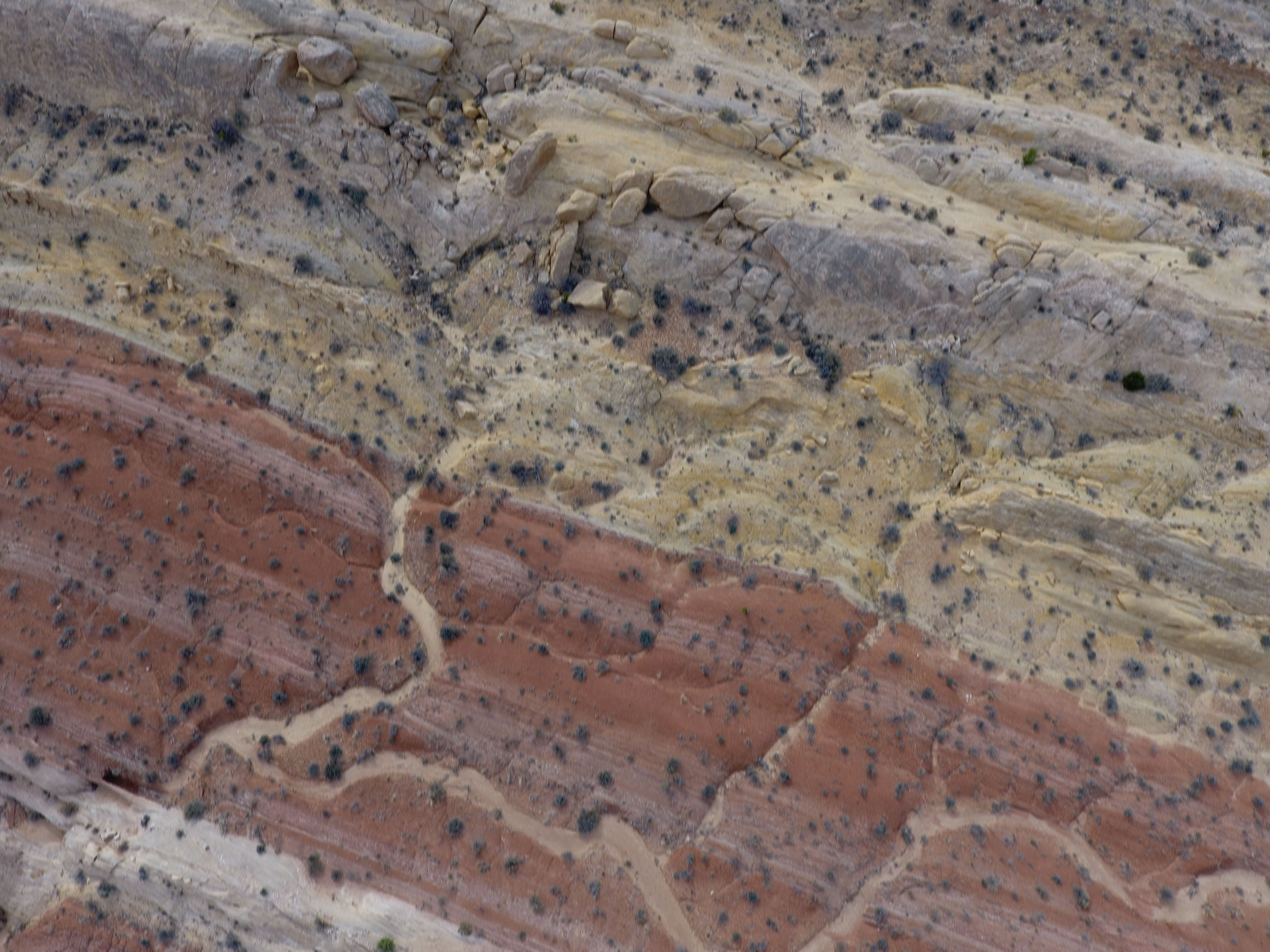}
  \caption{Vertical 10}
  \label{fig:vertical10}
\end{subfigure}%
\begin{subfigure}{0.5\textwidth}
  \centering
  \includegraphics[width=.8\linewidth]{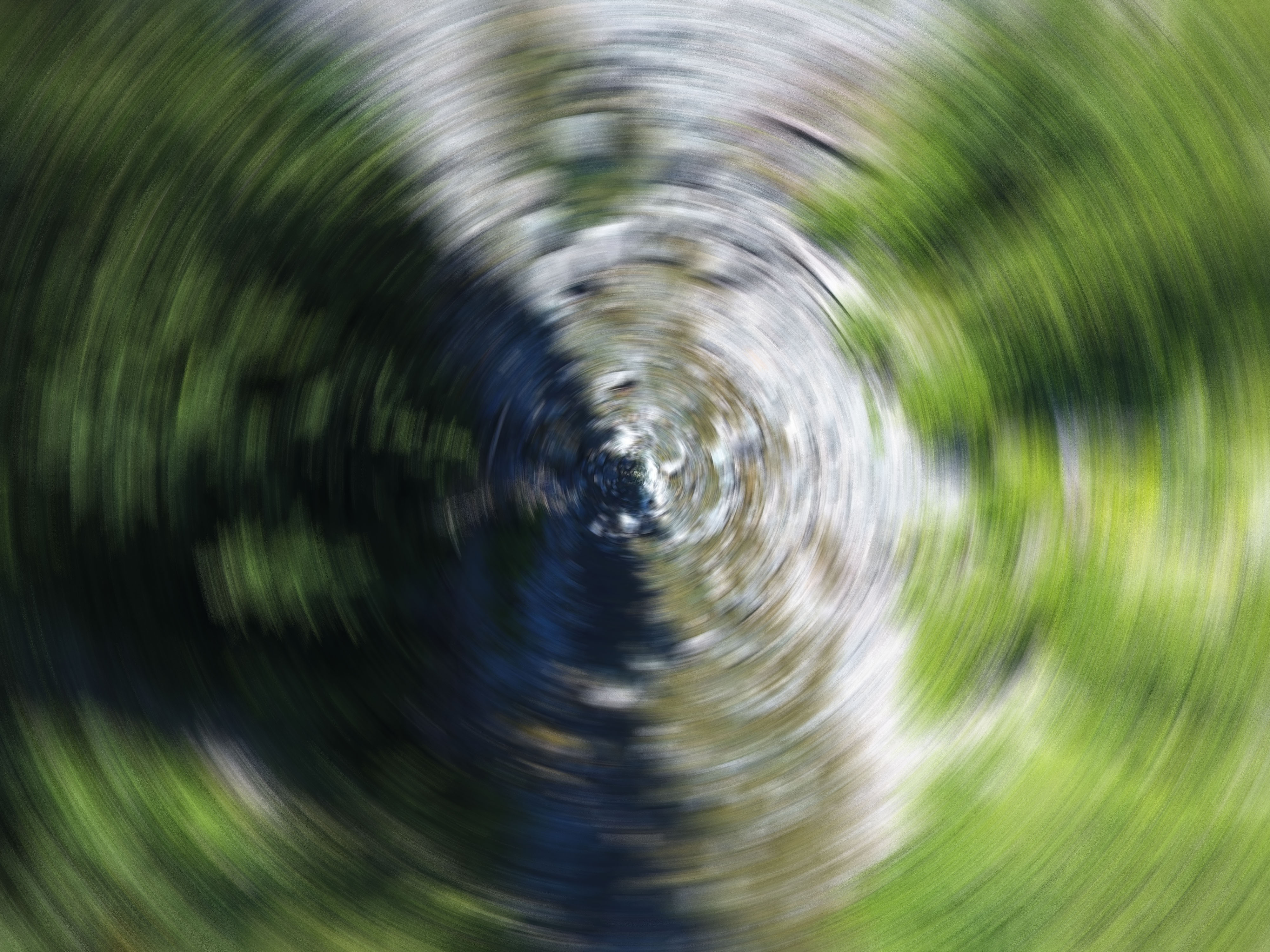}
  \caption{Spin 10}
  \label{fig:spin10}
\end{subfigure}
\caption{Generated vertical shift and radial spin blur images with shift extent 10 and spin extent 10. Vertical or horizontal shift blur comes from the constant movement of drones in geo-survey, and severe radial spin can be from strong wind, for example.}
\label{fig3}
\end{figure}

\begin{figure}
\centering
  \includegraphics[width=0.85\linewidth]{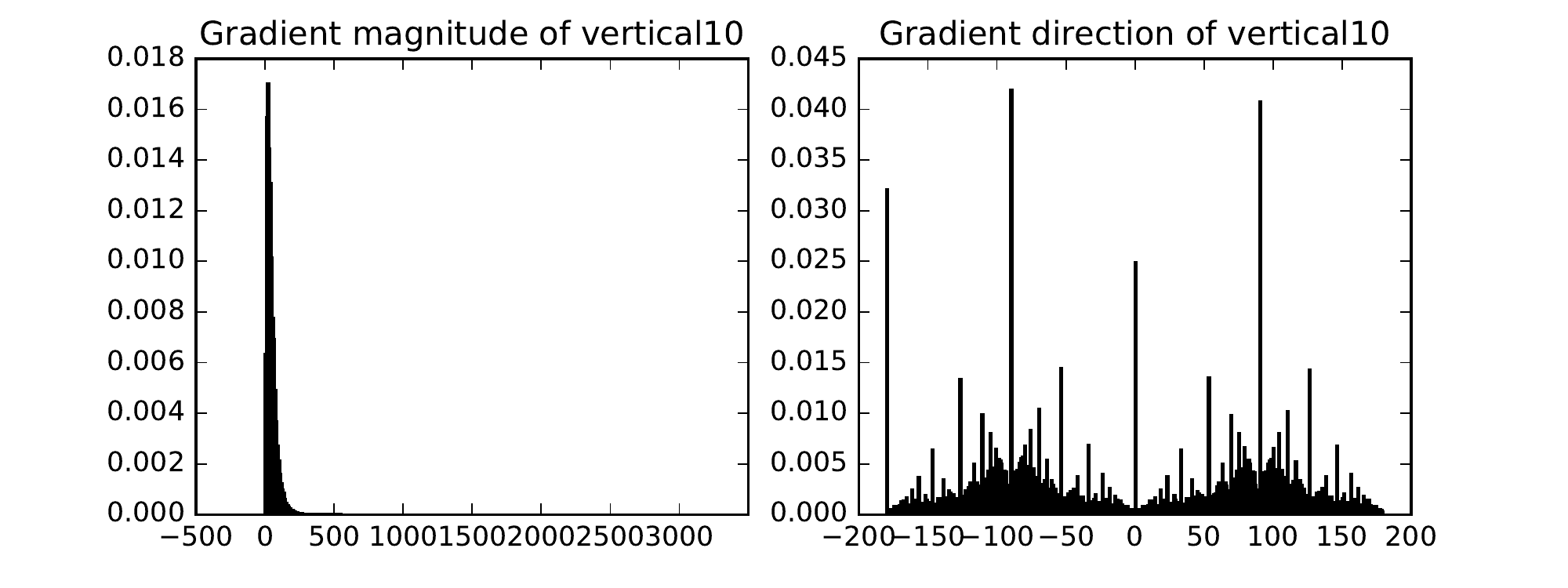}
  \caption{Gradient histograms for vertical blur 10. Clear coefficient $\alpha$ is 5e-07}
  \label{fig:vertical10gradient}
\end{figure}

\begin{figure}
  \centering
  \includegraphics[width=0.85\linewidth]{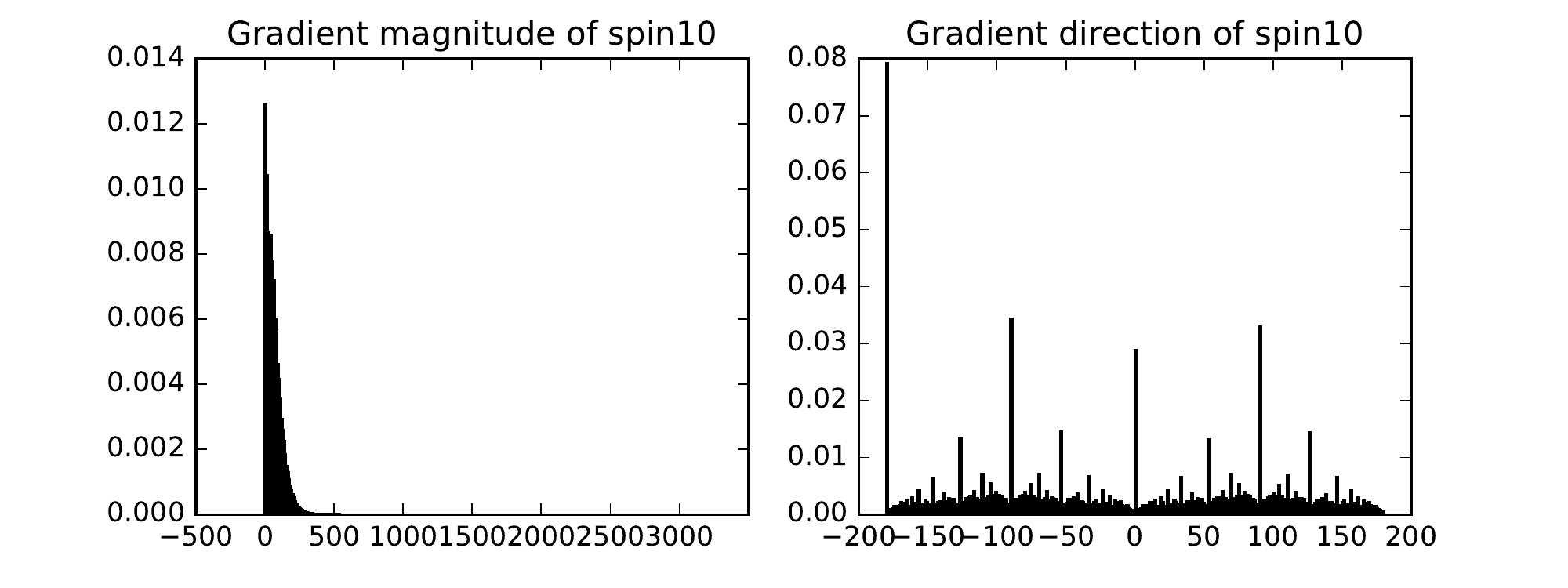}
  \caption{Gradient histogram for spin 10. Clear coefficient $\alpha$ is 0}
  \label{fig:spin10gradient}
\end{figure}

Figure \ref{fig3} shows two very blurry images, one from shifting vertically and one from spinning around the center of the image. We show our feature analysis results in Figure \ref{fig:vertical10gradient} and \ref{fig:spin10gradient}. Notice the gradient model ignores the difference between these two types of blur, and the gradient magnitude histogram simply shows the severity: even fewer pixels counts than histograms for images in Figure \ref{fig1} for larger gradient magnitude. Both gradient direction histograms look fuzzy, spin blur has more dispersed spikes and vertical blur has more taller spikes. Clear coefficient $\alpha$ for clear image is magnitudes higher than those for blurry images, and we can always adjust a preset threshold to normalize our output feature. 


\subsection{Feature analysis: singular value decomposition (SVD)}
The image gradient model is based on edge sharpness information. Inspired by \cite{svd}, we decompose an image array into multiple rank 1 matrices, call {\it eigen-images}, via the SVD decomposition. Each eigen-image correspond to a singular value: blurred images to larger singular values, and sharp images to smaller singular values. This is because blurry images suppress lose high-frequency information, which is reflected in less significant eigen-images. 

Consider a grayscale image $F$ of size $M\times N$, performing SVD on $F$ to get

\begin{equation}
    F=U\Sigma V^T.\label{svd}
\end{equation}
The column-row multiplication of $U\Sigma$ and $V^T$ allows writing $F$ as the sum of $r$ (the rank of $F$) rank 1 matrices:
\begin{equation}
    F=U\Sigma V^T=\sigma_1\mathbf{u}_1\mathbf{v}_1^T+...+\sigma_r\mathbf{u}_r\mathbf{v}_r^T,\label{svdDecomp}
\end{equation}
here $\mathbf{u}_i$, $\mathbf{v}_i$ denote columns of $U$ and $V$, $\sigma_i$ the singular values, arranged in descending order. Then image $F$ is written as a weighted sum of eigen-images $\mathbf{u}_i\mathbf{v}_i^T$, with weights $\sigma_i$. 

In image compression with SVD, we use low rank approximation with eigen-images: with a preset $k$, we obtain compressed image $F_k=\sigma_1\mathbf{u}_1\mathbf{v}_1^T+...+\sigma_k\mathbf{u}_k\mathbf{v}_k^T$ of $F$. The compression omits details of the image, which correspond to omitting smaller singular values and their corresponding eigen-images. Thus, the fewer small singular values we have, the less detail an image contains. See Figure \ref{figsvd}. We then define a SVD blur degree $\beta$, and propose the following algorithm for blur detection:

  \begin{figure}
\begin{subfigure}{.5\textwidth}
  \centering
  \includegraphics[width=0.75\linewidth]{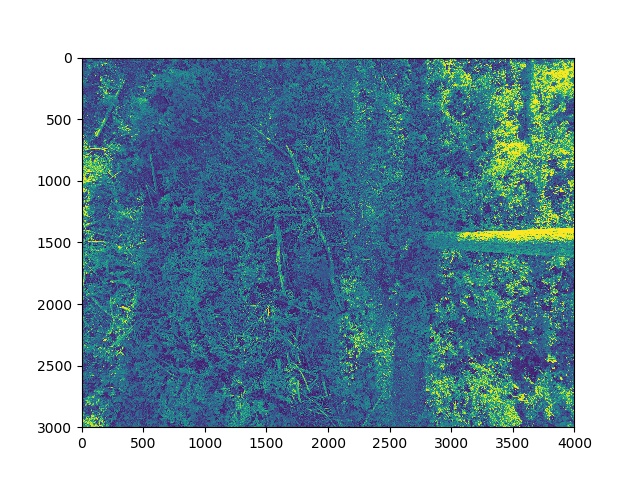}
  \caption{Reconstructed with all 3000 singular values}
  \label{fig:fullSVD}
\end{subfigure}%
\begin{subfigure}{.5\textwidth}
  \centering
  \includegraphics[width=0.75\linewidth]{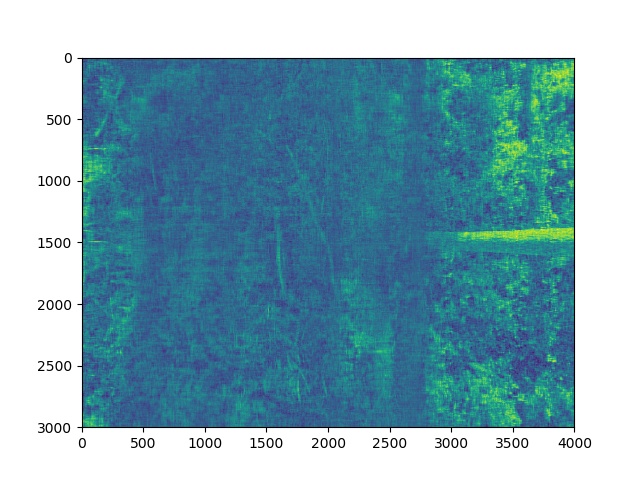}
  \caption{Reconstructed with the 50 biggest singular values}
  \label{fig:50SVD}
\end{subfigure}
\caption{Clear image reconstructed with all singular values versus the 50 biggest singular values. Details were lost with fewer singular values.}
\label{figsvd}
\end{figure}

\begin{framed}
\noindent {\bf Input}: Image $F$ of size $M\times N$

\noindent {\bf Output feature}: SVD blur degree $\beta$.

\begin{enumerate}
\item Compute SVD decomposition of the $M\times N$ image. Denote the singular values by $\{\sigma_i\}^n_{i=1}$
\item Compute blur degree: $\beta = \frac{\sum^k_{i=1}\sigma_i}{\sum^n_{i=1}\sigma_i}$, where $k$ is preset by user.
\item The blur of image is measured by the blur degree. The higher the degree is, the more blurry the image is.
\end{enumerate}
\end{framed}


    











We performed experiments with images from our geo-survey (sizes around 3000 by 4000), and have found when taking $k=300$, $\beta$ for blurry images were higher than 0.63, whereas for clear we get below 0.51; for $k=50$, $\beta$ for blurry images were all higher than 0.3, and lower than 0.2 for clear images. Independent of $k$ value, blurry and clear images have a distinct linear range for blur degree.

\begin{figure}[h]
\begin{subfigure}{.5\textwidth}
  \centering
  \includegraphics[width=0.7\linewidth]{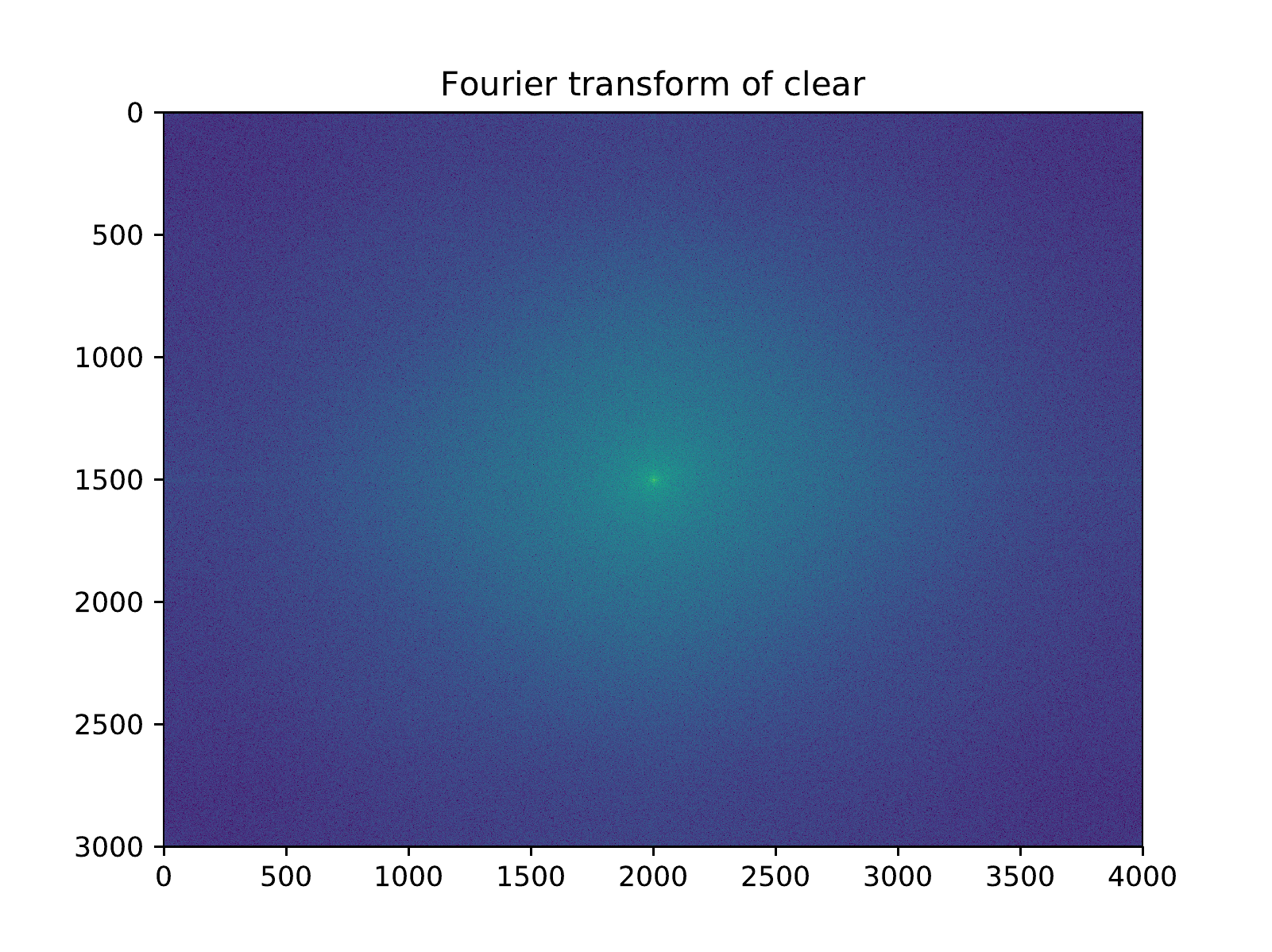}
  \caption{FFT of Clear image}
  \label{fig:clearfft}
\end{subfigure}%
\begin{subfigure}{.5\textwidth}
  \centering
  \includegraphics[width=0.7\linewidth]{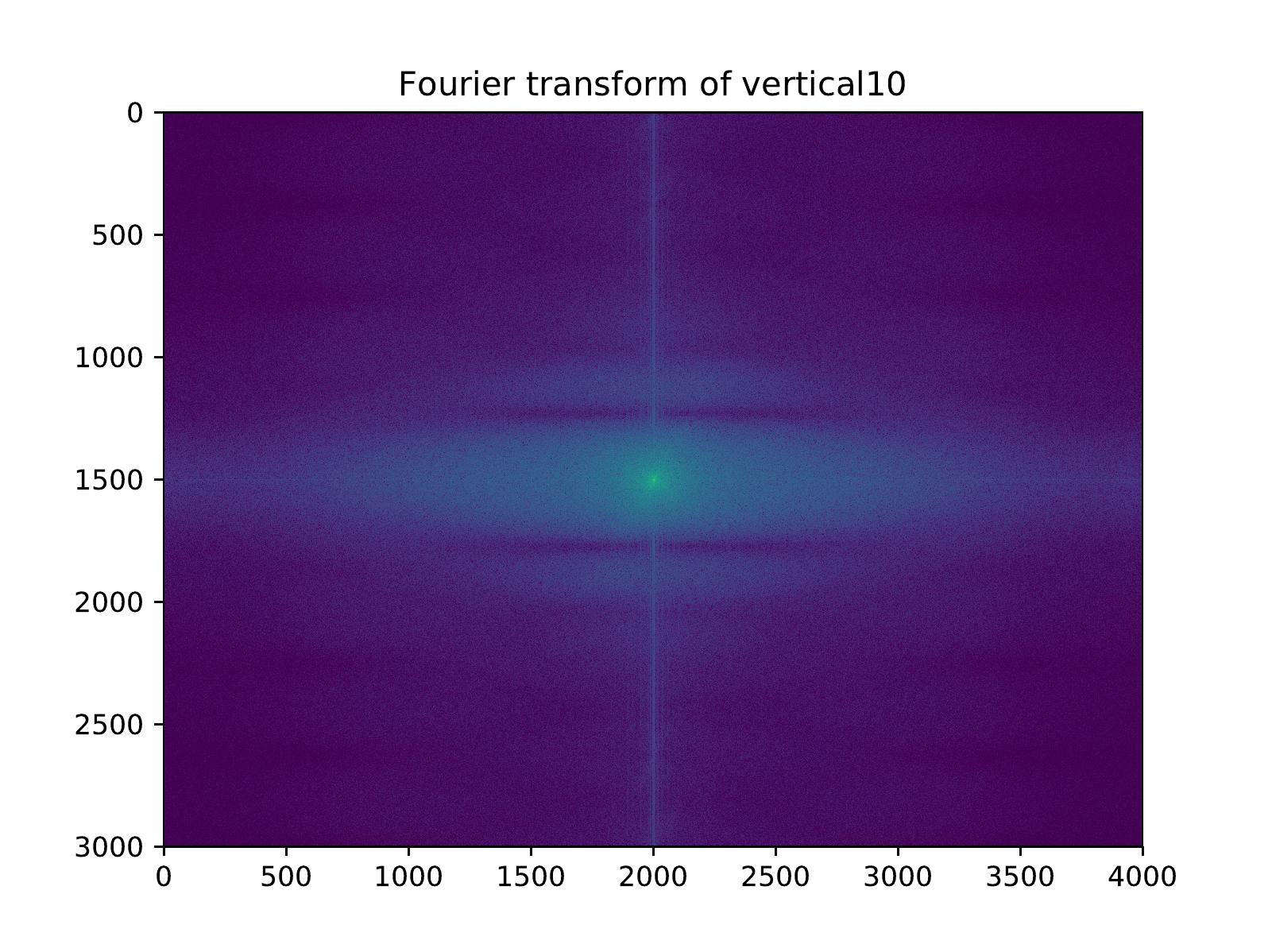}
  \caption{FFT for vertical 10 image}
  \label{fig:spin1fft}
\end{subfigure}%


\begin{subfigure}{.5\textwidth}
  \centering
  \includegraphics[width=0.7\linewidth]{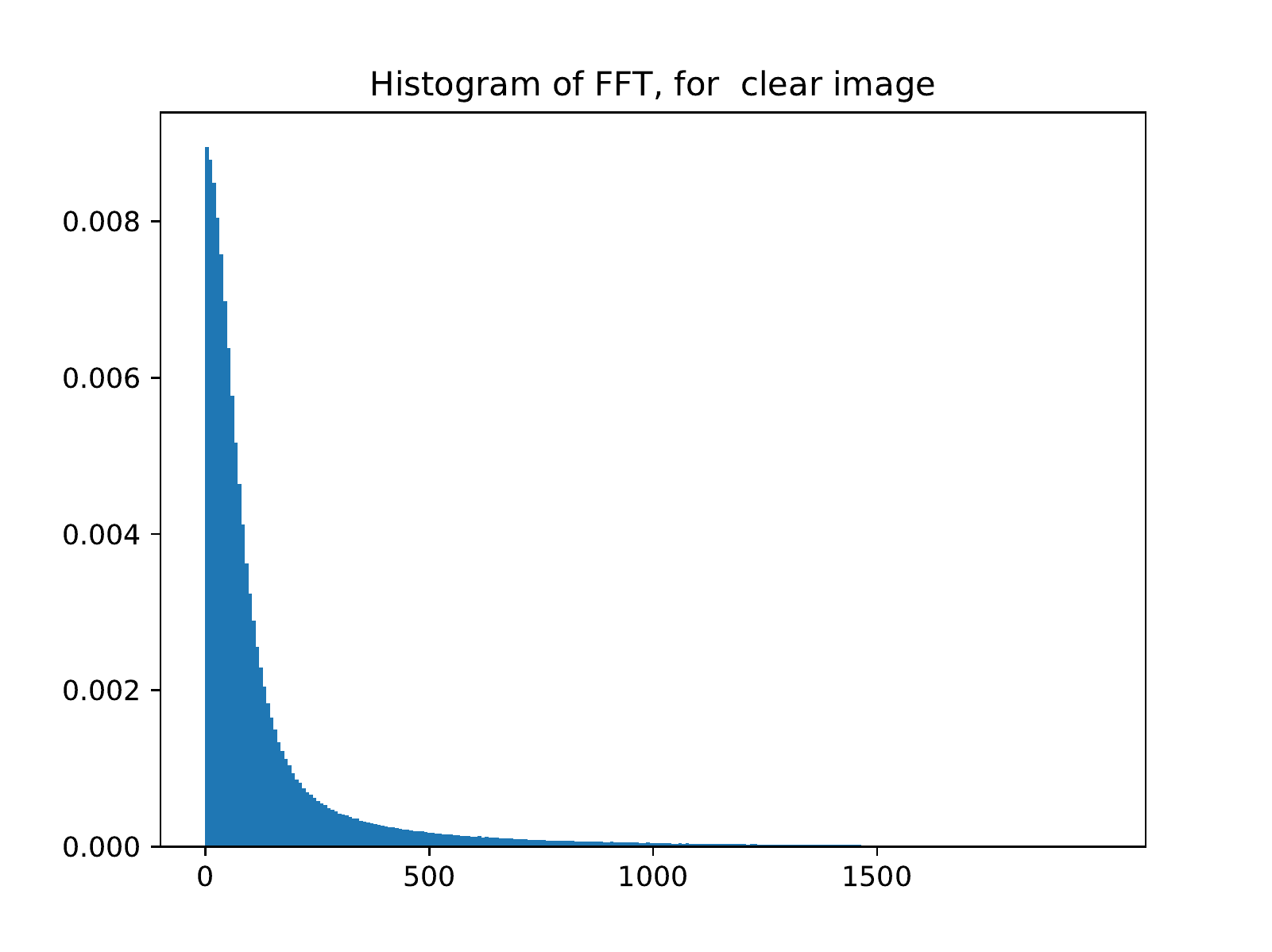}
  \caption{Histogram of FFT of clear image}
  \label{fig:clearhistogram}
\end{subfigure}
\begin{subfigure}{.5\textwidth}
  \centering
  \includegraphics[width=0.7\linewidth]{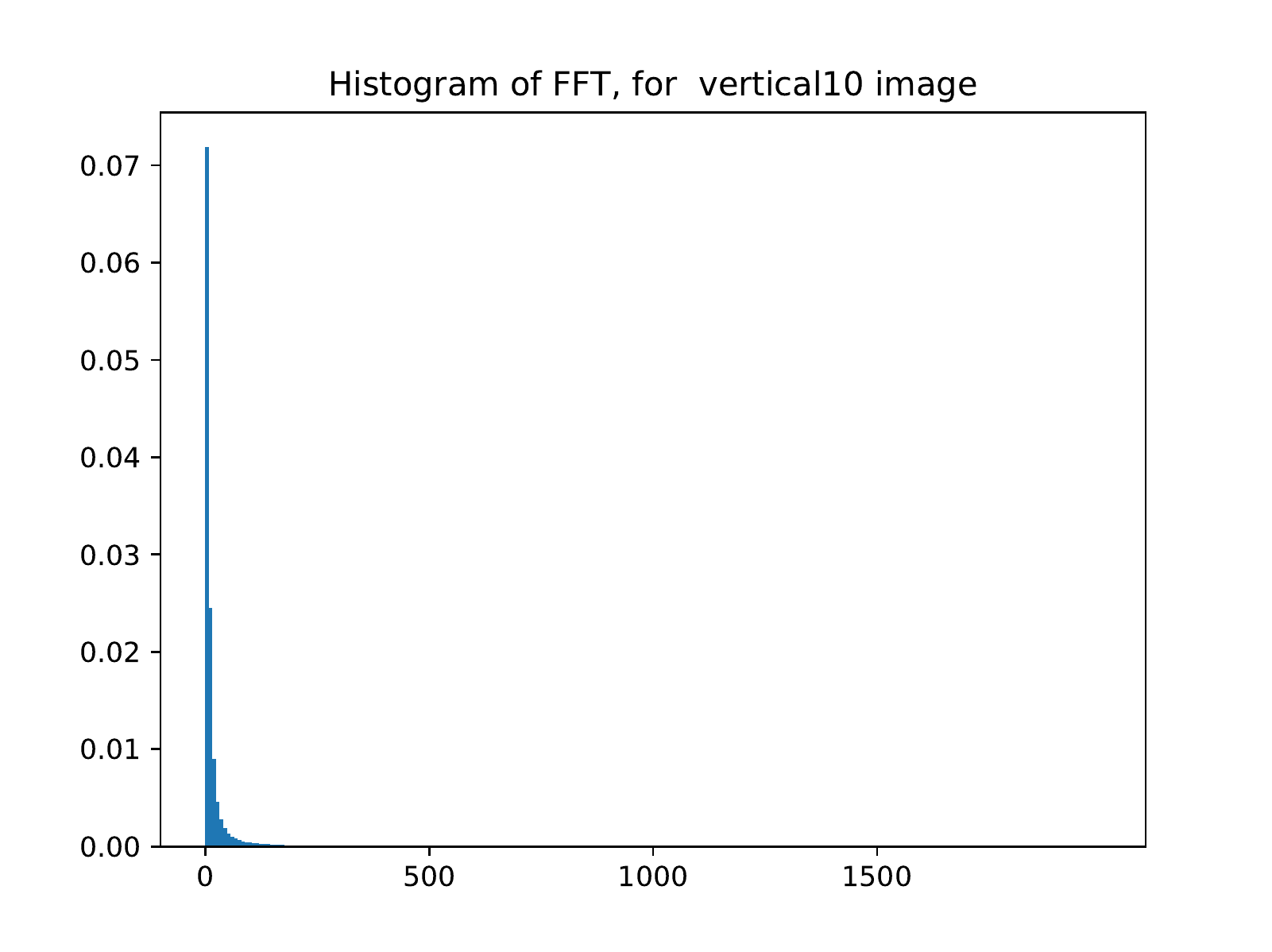}
  \caption{Histogram of FFT for vertical 10}
  \label{fig:spin1ffthistogram}
\end{subfigure}
\caption{2D FFT of clear image and vertical 10 image. More higher frequency components in the center in clear image. In the histogram, we see more higher frequencies in clear image.}
\label{fig_fft_spin1}
\end{figure}


\subsection{Feature analysis: Fourier Transform and frequency analysis}
An image is an array of data containing information in the spatial domain. Fourier transform of an image contains information in the frequency domain, which means the rate of change of intensity per pixel. In a region of a grayscale picture, if the change from white to black (or vice verse) takes many pixels, then the change is slow and corresponds to low frequency; if the process only takes a few pixels, then the intensity varies fast, corresponding to high frequency. Sharper image has more high frequency components, but blurry image has less high frequency components. We use this characteristic of images to train the machine to distinguish clear and blurry images. 

Fourier transform (FT) is an efficient computational method for image processing, and we use FT histogram vector as training feature \cite{de_masilamani}:

\begin{framed}
\noindent {\bf Input}: Image $F$ of size $M\times N$

\noindent {\bf Output feature}: FT histogram vector for image and clear estimate $\gamma$.

\begin{enumerate}
\item Compute the shifted FT of image as a $M\times N$ array, then convert the $M\times N$ array to a vector of length $MN$. Denote by $\mathbf{s}$.
\item find the maximum of the centered FT component: $a = max(abs(\mathbf{s}))$
\item Count the total number $T_H$ of pixels in image whose pixel value $>\ threshold$ (specified by user). \cite{de_masilamani} takes $threshold=a/1000$
\item compute percentage of high quality pixels of all pixels, define to be the clear estimate $\gamma$.
\end{enumerate}
\end{framed}


Figure \ref{fig_fft_spin1} shows the 2D FFT and corresponding histograms for clear and vertical 10 images. The 2D FFT figures show that clear images have more big frequencies in the center, and blurry images have fewer. Moreover, the FFT of vertical shift blur images shows the shift and direction of the shift. As expected, normalized clear estimate $\gamma$ is the biggest for clear image (0.0427), and as spin extent gets bigger, the clear estimate gets smaller: spin 1 is 0.0075, and spin 10 is 0.0017. Although FFT histogram ``thinks" vertical blur is clearer than spin blur with a clear coefficient of 0.0032, clear estimates of all blur images are one magnitude smaller than that of the clear image. We thus conclude this is a valid feature for blur detection in our data. 

\section{Future work}
With all three feature formulations, our experiments demonstrate that they are effective in distinguishing blurry images from clear ones. Image gradient model is image independent, although it depends on the geometric information of the image. This could be especially useful when studying image from Geoscience. SVD blur detection decomposes images into different levels of details, and measures blur by the percentage of singular values corresponding to non-detailed information. FFT studies the frequency of pixels in an image: sharper images corresponds to larger frequencies, and more larger frequencies implies clearer images. 

In the future, we will incorporate more machine learning techniques to detect blurry images in a large library of HD geo-survey image data with the above feature models. We have already done experiments with support vector machine (SVM), and are working on implementation with neural network (NN) and deep learning technoques. 


%
%
%
%

\end{document}